\documentclass[review]{elsarticle}

\usepackage{lineno,hyperref}
\modulolinenumbers[5]

\journal{International Journal of Multiphase Flow}

\usepackage{color}
\usepackage{multirow}
\usepackage{dcolumn}
\usepackage{multicol}
\usepackage{graphicx}
\usepackage{caption}
\usepackage{amssymb}
\usepackage{epstopdf}
\usepackage{mathtools}
\usepackage{float}
\usepackage{bm}

\usepackage{xcolor,etoolbox}

\biboptions{authoryear}

\biboptions{sort&compress}

\usepackage{color}

\newcommand{\ks}{\textcolor{black}} 

\begin{document}

\begin{frontmatter}

\title{An investigation on the impact of two vertically aligned drops on a liquid surface}
\author{Akash Paul$^a$, Bahni Ray$^b$, Kirti Chandra Sahu$^c$ and Gautam Biswas$^{a}$\footnote{gtm@iitk.ac.in} }
\address{$^a$Department of Mechanical Engineering, Indian Institute of Technology Kanpur, Kanpur - 208016, Uttar Pradesh, India  \\ 
$^b$Department of Mechanical Engineering, Indian Institute of Technology Delhi, Hauz Khas, New Delhi-110 016, India \\
$^c$Department of Chemical Engineering, Indian Institute of Technology Hyderabad, Sangareddy 502 284, Telangana, India}

\begin{abstract}
The dynamics of two vertically coalescing drops and a pool of the same liquid have been investigated using a Coupled Level Set and Volume of Fluid (CLSVOF) method. Such a configuration enables us to study the dynamic interaction of an arbitrary-shaped liquid conglomerate, formed owing to drop-drop coalescence, with a pool. Similar to drop-pool and drop-drop interactions, partial coalescence is observed when a conglomerate interacts with a pool. The presence of the pool below the father drop is found to influence the coalescence characteristic of the two drops. At the same time, the movement of the capillary waves resulting from the interaction of two drops governs the coalescence dynamics of the conglomerate with the pool. As liquid interfaces interact and generate capillary waves at multiple locations, complex trajectories of capillary waves are observed, which play a crucial role in determining the pinch-off characteristics of the satellite during conglomerate-pool interaction. We examine the effect of the ratio of the diameters of the lower/father drop to the upper/mother drop ($D_r$) on the coalescence dynamics while maintaining the size of the mother drop constant. The variation in the coalescence dynamics due to change in $D_r$ is quantified in terms of the residence time ($\tau_r$), pinch-off time ($\tau_p$) and the satellite diameter to conglomerate diameter ratio ($D_s/D_c$). The coalescence dynamics of the conglomerate is then compared with that of an equivalent spherical drop of the same volume and also with that of a drop initialized with the same shape as that of the conglomerate. Finally, the regions of complete and partial coalescence for the conglomerate-pool interactions are demarcated on the Weber number - diameter ratio ($We-D_r$) space.
\end{abstract}

\end{frontmatter}


\section{Introduction}\label{sec:introduction}
Coalescence of a drop on a liquid pool has intrigued researchers for centuries due to its relevance to natural phenomena \citep{thomson1886v,pumphrey1989underwater} and industrial applications \citep{stone2004engineering,thoroddsen2008high}. The behaviour of a drop interacting with liquid or solid surfaces is highly dependent on the momentum with which it interacts with the surface \citep{ajaev2021levitation}. A drop impacting a liquid-air interface involves complex physics in addition to a wide range of practical applications. This phenomenon includes the generation of capillary waves that cause satellite droplets to pinch off from the free surface (a phenomenon known as partial coalescence), crater formation (at low impact velocity), and splashing (at high impact velocity).

An impinging drop in a liquid pool either coalesces completely or partially at low impact velocities, while it may cause splashing if the impact velocity is high. In a certain parametric range, the drop may also bounce back. If the momentum is small, the drop may float on a thin layer of the ambient fluid trapped between the drop and the liquid surface for a finite time (residence time) before getting involved in the coalescence process \citep{reynolds1881floating,deka2019coalescencejfm,kirar2020coalescence}. During this time, the trapped air is released, and the droplet sinks into the pool. Researchers have also attempted to estimate the residence time of a drop on a liquid surface when it floats on the free surface due to the air cushion that exists between the drop and the pool \citep{charles1960mechanism,kirar2022coalescence}. Subsequently, a hole is formed in the film, and the drop experiences a partial coalescence. When the drop comes in contact with the liquid surface, the capillary waves are generated at the contact point and propagate up the drop. The capillary pull from the top and the negative horizontal momentum at the neck cause pinch off, and a satellite drop is formed, with a radius less than that of the falling drop \citep{chen2006partial,blanchette2006partial,ray2010generation}. Unlike complete coalescence, a satellite drop pinches off for partial coalescence, producing a cascade of self-similar events that repeat until the final drop completely merges with the liquid pool. This repeated occurrence of the secondary drops from the parent drop is called a coalescence cascade \citep{thoroddsen2000coalescence}. \citet{honey2006astonishing} observed up to six secondary drops in which the same fraction reduces the drop size in each cycle of partial coalescence. They attributed this primarily to the capillary force at the pinch-off. \citet{gilet2007critical} investigated the ratio between the secondary and primary drops and the role of the capillary waves on partial coalescence criteria. The partial coalescence is a consequence of the competition between the horizontal and vertical collapse rates. The horizontal collapse is always confined to regions lower than the equator of the initial drop. Hence the partial coalescence is ensured \citep{gilet2007critical}. As the impact velocity or size of the primary drop increases, it undergoes complete coalescence, followed by a splashing phenomenon. A few researchers have also investigated the effect of the temperature difference between the drop and pool on the coalescence dynamics and residence time of droplets on liquid surface  \citep{shen2022durably,kirar2020coalescence,davanlou2016role,thrivikraman2021confined,shahriari2017electrostatic}.

Another common yet fascinating phenomenon is the dynamics of the coalescence of two drops \citep{gilet2007controlling,yoon2007coalescence}. \citet{thoroddsen2005coalescence} studied the coalescence of both equal and unequal-sized pendent and sessile drops and formulated a model to estimate coalescence speed. Similar to drop-pool interactions, complete and partial coalescence can also be observed in two-drop configurations. \citet{thoroddsen2005coalescence} reported two different types of coalescence phenomena: (i) first-stage coalescence where the mother drop after coalescing with the father drop produces a satellite drop by draining some liquid on the father drop, and (ii) second-stage coalescence where the mother drop, after coalescing with the father drop, although reaches the incipience of pinch-off, the neck expands without the detachment of the satellite drop. The undetached secondary drop then coalesces with the father drop and forms a column structure again and starts the necking which finally leads to the pinch-off of a satellite drop. This is called second-stage coalescence because the satellite is formed after a two-stage coalescence event, the first necking without pinch-off and the second necking that leads to pinch-off. Based on experiments with water and water-glycerin mixture in an air matrix, \citet{zhang2009satellite} reported that the minimum size ratio of the two drops below which satellite formation inhibits is 1.55. \citet{deka2019coalescence} numerically investigated the interaction of two unequal-sized drops that provides important insights into the transition from complete to partial coalescence and also elaborately studied the effects of the governing dimensionless parameters on the coalescence dynamics. A similar configuration was investigated by \citet{singh2022dynamics}. \citet{hassanzadeh2019numerical} numerically investigated the head-on collision of two drops in a vertical channel and studied the resulting oscillations after merging the drops. Irrespective of whether the drop interacts with a pool or another drop, it was found that increasing the viscosity of the liquid slows down the coalescence process \citep{ray2010generation,thoroddsen2005coalescence}. The effect of the physical properties of the fluid and external electric field on the coalescence dynamics of drops have been discussed in great detail in \cite{kavehpour2015coalescence}.

\ks{In droplet-film interaction, \citet{tang2016} illustrated that increasing the liquid film thickness below a critical inertial limit shows the non-monotonic transition between bouncing and merging. \citet{Tang2019} investigated the dynamics of the gas layer when a drop bounces after impacting liquid films of various thicknesses at different Weber numbers. Furthermore, different shapes of the interfacial gas layer have been observed for higher film thickness during the approaching and bouncing stages of the drop. \citet{saha_wei_tang_law_2019} experimentally investigated the evolution and motion of the vortex ring generated during the coalescence of a drop in a pool of the same liquid. They also classified the effects of inertial, capillary and viscous forces in determining the motion of the vortex ring. In droplet-droplet interaction, collisions may exhibit various outcomes. \citet{qian_law_1997} identified five regimes (coalescence after minor deformation, bouncing,  coalescence after substantial deformation, coalescence followed by separation for near head-on collisions, and coalescence followed by separation for off-centre collisions) while conducting experiments with water and hydrocarbon droplets. They also suggested that coalescence is facilitated when the surrounding medium contains vapour of the interacting liquid masses. \citet{tang2012} identified distinct outcomes of the head-on collisions of two droplets and discussed the variation of the critical Weber number that separates bouncing and coalescence regimes with the size ratio of the two colliding drops. \citet{Kai_PRA_2015} numerically investigated the collision dynamics and internal mixing of droplets of non-Newtonian fluids. They simulated droplet collisions for different combinations of shear-thickening and shear-thinning properties and identified the outcomes in coalescence and mixing dynamics.}

Droplet-based microfluidic devices have become increasingly popular in recent years and are employed in many different biochemical and molecular biological tests \citep{sanchez2019recent,anna2016droplets}. Moreover, the collision of droplets is also relevant in atomization and spray combustion processes \citep{cong2020numerical}. A microscopic liquid bridge forms when two liquid drops come in contact and quickly spreads as the two drops coalesce to form one. The coalescence of two or more drops with another interface is another circumstance that frequently occurs in many applications. \citet{paulsen2014coalescence} examined the effect of the viscosity of the outer fluid on the coalescence dynamics of the two drops. \citet{sprittles2012coalescence} numerically investigated the coalescence of two identical liquid drops using different models and compared them with the experiments. They found that the interface formation model provides a more accurate description of the natural coalescence process. \citet{chan2011film} reviewed different experimental studies on the interaction and coalescence of deformable drops and bubbles and provided detailed quantitative information on the spatial and temporal evolution of interfacial deformations. \citet{damak2018expansion} experimentally studied the interaction and agglomeration of two droplets on a superhydrophobic surface. They identified inertial-capillary and viscous regimes and described both the expansion and retraction phases, and established generalised models to characterise the distinct phenomena observed in their experiments. It was observed that the influence of various drop sizes produces drastically different behaviours while impacting a surface \citep{LiuNatCommun}. They found that the Kelvin-Helmholtz instability promotes splash and that the quantity of splash depends on how much energy is lost before the liquid takes off. \citet{he2021drop} investigated the impact of a drop on an ultra-thin film and found that the drop exhibits contact bouncing behaviour at the same impact velocity for both low and high film thicknesses. It displays a range of behaviours for intermediate film thickness and when the impact velocity is varied. They presented a phase diagram based on these observations. Despite the significance in practical applications, only a few researchers have examined the effects of multiple droplets (organised in parallel or in a line) on surfaces \citep{poureslami2023simultaneous,markt2021high,benther2021heat}. Recently, \citet{poureslami2023simultaneous} investigated the effect of simultaneous double droplets on a pool of molten phase change material (PCM), which results in the droplets evaporating and the PCM solidifying. \citet{benther2021heat} considered the dynamics of multiple droplets in different configurations, such as two droplets falling in-line and side-by-side, train of droplets, spray of droplets, etc.

As the above-mentioned review indicates, although the drop-pool and drop-drop interactions have been studied extensively, to the best of our knowledge, the paradigm where these two processes occur subsequently or concurrently has not been investigated yet, which has many consequences in practical applications, such as ink-jet printing and electronic cooling \citep{stone2004engineering,thoroddsen2008high}. Therefore, in the present study, we intend to understand the complex phenomenon of the interactions between a conglomerate formed by two coalescing drops and a pool of the same liquid. We examine the impact of the ratio of the diameter of the father ($D_f$) to mother ($D_m$) drops and the Weber number ($We$) on the coalescence dynamics of the conglomeration of two drops of different sizes in a liquid pool while maintaining the size of the mother drop fixed. \ks{Our study provides an insightful interpretation of the
physical processes involved during the impact of two vertically aligned drops on a liquid surface and attempts to provide with meaningful explanations for the observed behaviour.}

The rest of the paper is organized as follows. The problem is formulated in $\S$\ref{sec:formulation}, the numerical method and validation of the present solver are discussed in $\S$\ref{sec:num}. The results are presented in $\S$\ref{sec:results} and concluding remarks are summarized in $\S$\ref{sec:conclusion}.

\section{Formulation}\label{sec:formulation}

\begin{figure}[h]
\centering
\includegraphics[width=0.45\textwidth]{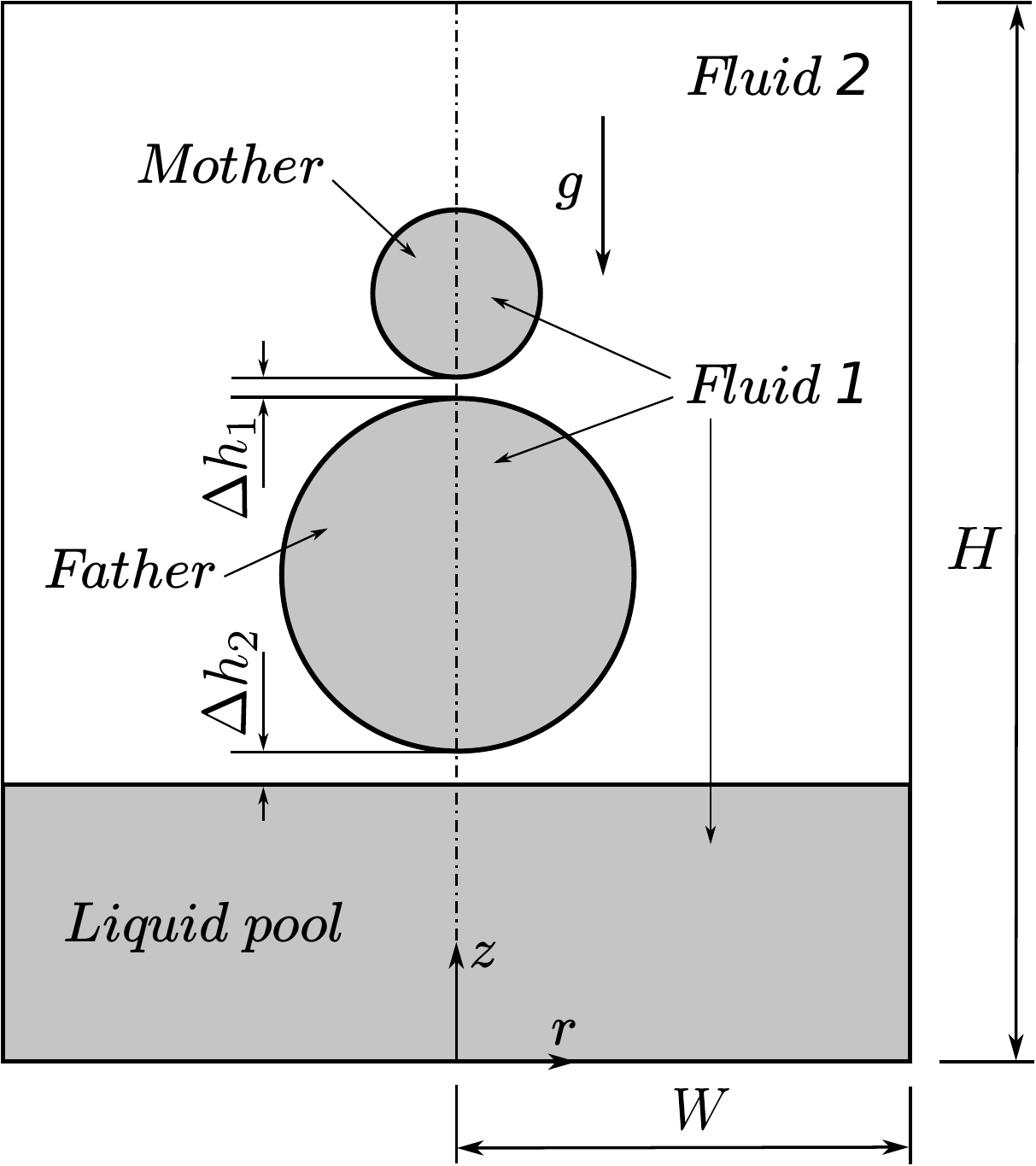}
\caption{Schematic diagram of the computational domain.}
\label{fig:fig1}
\end{figure}

We perform numerical simulations to investigate the coalescence dynamics of conglomerated drops in a liquid pool using a Coupled Level Set and Volume of Fluid (CLSVOF) method. The schematic diagram of the computational domain is shown in figure \ref{fig:fig1}. An axisymmetric cylindrical coordinate system $(r,z)$ with its origin at the centre of the bottom wall (as shown in figure \ref{fig:fig1}) is incorporated in our numerical simulations. Both the drops are initialized in a vertically in-line configuration above the pool. The separation distances between the two drops $(\Delta h_1)$, and the father drop and pool $(\Delta h_2)$ are kept very small ($\mathcal{O}(0.01D_m)$, and $\Delta h_2/\Delta h_1 = 1.75$ (unless explicitly mentioned). The positioning of the drops and the pool is done in such a way that the mother and father drops begin coalescing with one another before the conglomerate interacts with the pool. A small initial velocity ($\thickapprox$ 0.05 ms$^{-1}$) is imposed on the mother drop for the cases with $We=0.11$. The father drop is initialized at rest.  The drops and the pool are of the same fluid ($fluid \ 1$), and they coalesce in the surrounding medium ($fluid \ 2$). Depending on the diameter ratio of the father and mother drops $(D_f/D_m)$, the width $(W)$ and height of the domain $(H)$ are varied in the ranges $3D_m$ to $9D_m$ and $6D_m$ to $18D_m$, respectively.

\subsection{Governing equations}

The conservation equations for mass and momentum for incompressible Newtonian fluids are given by
\begin{eqnarray}
\nabla \cdot U = 0, \label{eqn:eq1}
\end{eqnarray}
\begin{eqnarray}
\rho (\phi) \left(\frac{\partial \bm{U}}{\partial t} + \bm{U}. \nabla \bm{U} \right) = -\nabla P + \nabla \cdot [\mu(\phi)(\nabla \bm{U} + \nabla \bm{U}^T)] + \sigma \kappa \bm{n} \delta_s + \rho(\phi)\bm{g}.   \label{eqn:eq2} 
\end{eqnarray}
Here, $\bm{U} = (u,v)$ is the velocity vector having $u$ and $v$ as the radial and axial components of velocity, respectively. $P$ denotes the pressure field, $\bm{g}$ is the gravitational field, $\sigma$ represents surface tension, $\kappa$ is the mean curvature of the interface, $\bm{n}$ is the unit normal vector on the interface boundary, and $\delta_{s}$ is the interface delta function which is zero elsewhere except on the interface. In Eq. (\ref{eqn:eq2}), the surface tension force is included as a body force term using the formulation proposed by \cite{brackbill1992continuum}. The smoothness of the interface is attained using a level set function $\phi$, which is expressed as the signed distance from the interface as
\begin{equation}
 \phi =
    \begin{cases}
     -d, & \text{in the fluid 2 region,}\\
     0, & \text{at the interface,}\\
     +d, & \text{in the fluid 1 region.}
    \end{cases}       
\end{equation}
The interfacial dynamics is computed by solving the advection equations for the volume fraction $F$ and level set function $\phi$, which are given by
\begin{equation}
    \frac{\partial F}{\partial t} + \bm{U} \cdot \nabla F = 0, \label{eqn:eq4} \\ 
\end{equation}
\begin{equation}
    \frac{\partial \phi}{\partial t} + \bm{U} \cdot \nabla \phi = 0. \label{eqn:eq5}
\end{equation}
The normal and the curvature are calculated from the level set
function $\phi$ as
\begin{eqnarray}
    \bm{n} &=& \frac{\nabla \phi}{|\nabla \phi|},\\
    \kappa &=& -\nabla \cdot \bm{n}  = -\nabla \cdot \frac{\nabla \phi}{|\nabla \phi|}.
\end{eqnarray}
A smoothed Heaviside function is defined based on the level set function as
\begin{equation}
 F = H(\phi) =
    \begin{cases}
     1, & {\rm if}  \phi > \delta,\\
     \frac{1}{2}+\frac{\phi}{2\delta}+\frac{1}{2\pi}\left[sin(\frac{\pi\phi}{\delta})\right], & {\rm if} \ \phi \leq \delta,\\
     0, & {\rm if} \ \phi < -\delta.
    \end{cases}       
\end{equation}
Here, $\delta$ is the numerical thickness of the interface. It is the distance over which the phase properties are interpolated. The density, $\rho(\phi)$, and the dynamic viscosity, $\mu(\phi)$, of the medium are calculated using the smoothed Heaviside function as
\begin{eqnarray}
    \rho(\phi) &=& \rho_{1}H(\phi) + \rho_{2}(1-H(\phi)),  \\
    \mu(\phi) &=& \mu_{1}H(\phi) + \mu_{2}(1-H(\phi)).
\end{eqnarray}
Next, we discuss the numerical method used in our study.

\section{Numerical method}\label{sec:num}
The marker and cell (MAC) algorithm is used to solve the single set of governing equations on a staggered grid arrangement \citep{harlow1965numerical}. In such a grid arrangement, the scalar quantities are defined at the cell centers and the vector quantities, such as the components of the velocity are defined at the center of the cell faces to which they are normal. The governing equations are discretized using the finite-difference method. The grid size in the radial and axial directions is considered to be the same, i.e., $\Delta r = \Delta z$. The discretized  momentum equation (Eq. \ref{eqn:eq2}) is given by
\begin{eqnarray}
U^{n+1} &=& U^n + \left [-\nabla \cdot (U^n U^n) \right]\Delta t + g \Delta t \nonumber \\ 
&+& \left[\frac{-\nabla P^{n+1} + \nabla \cdot 2\mu(\phi^n)\bm{D}_v^n + \sigma\kappa(\phi^n)\bm{n}(\phi^n)} {\rho(\phi^n)}\right]\Delta t, 
\end{eqnarray}
where $\Delta t$ is the time-step and the $D_v = \frac{1}{2} (\nabla U + (\nabla U)^T)$ is the deformation tensor. The convective terms in the momentum equation are discretized using the higher order essentially non-oscillatory (ENO) scheme as described by \cite{chang1996level}, and the remaining space derivatives are discretized using the central difference scheme. The discretized form of the momentum equation is solved explicitly for the known volume fraction field $F^n$, which gives rise to the provisional velocity field. Such a velocity field may not be divergence-free since it does not satisfy the continuity equation in each cell. The compliance of the continuity equation is achieved by solving the corresponding pressure correction equation using the HYPRE multi-grid solver. Thus, after having achieved a divergence-free velocity field, the converged solution is obtained at a new time level.
Using this new velocity field, the advection equations of the volume fraction and level set function are solved to obtain the new volume fraction field $F^{n+1}$ and the level set function $\phi^{n+1}$, which are essential for interface reconstruction. The second-order conservative operator split advection scheme \citep{puckett1997high} is used for the discretization of the volume fraction advection equation (Eq. \ref{eqn:eq4}). In order to obtain higher accuracy, divergence correction is implemented \citep{gerlach2006comparison,puckett1997high,rider1998reconstructing} Thus, Eq. (\ref{eqn:eq4}) is reformulated into the conservative form along with the implementation of divergence correction as $\partial F/ \partial t + \nabla \cdot (F\bm{U}) = F\nabla \cdot \bm{U}$, which is then solved using the operator split advection scheme. The conservation of F is maintained by employing an implicit scheme in the first sweeping direction and an explicit scheme in the second direction as suggested by \cite{puckett1997high}. The approach is made second-order accurate by alternating the sweep directions in each time step \citep{rudman1997volume}. The level set advection equation (Eq. \ref{eqn:eq5}) is simultaneously solved in the corresponding directions by discretizing the convective terms using the ENO scheme. At each time step, after finding the updated volume fraction $F^{n+1}$ and level set function $\phi^{n+1}$, the level set function is reinitialized to the exact signed normal distance from the reconstructed interface by coupling level set function with volume fraction \citep{sussman2000coupled,son2002coupled,son2003efficient}. In the present work, the time-stepping procedure is based on an explicit method to maintain the stability of the solution. During the computations, the time steps are chosen to satisfy Courant-Friedrichs-Lewy (CFL), viscous, and capillary time conditions. 

In our simulations, the symmetry boundary condition is used at $r = 0$ and the free-slip boundary condition is employed at the side boundary ($r = W$). The no-slip and no-penetration boundary conditions are used at the bottom wall ($z = 0$), and the Neumann boundary condition is used at the top of the computational domain ($z = H$). Detailed formulation of the boundary condition used here can be found in  \cite{ray2010generation,deka2019coalescence}. \ks{It is worth mentioning that a noticeable limitation of the current study is the assumption of axisymmetry in the problem. In actual situations, this assumption may not be entirely valid, given that the shape of the conglomerate may not always exhibit symmetry about the vertical axis.}

The coalescence behavior of a drop is governed by interfacial tension, gravity, and viscosity of the drop and surrounding fluid. The relevant non-dimensional numbers are the Bond number, $Bo = \rho_c g D_m^2/\sigma$, Ohnesorge numbers, $Oh_1 = \mu_1/\sqrt{\rho_a \sigma D_m}$ and $Oh_2 = \mu_2/\sqrt{\rho_a \sigma D_m}$, Atwood number, $A = \rho_c/2\rho_a$, Weber number, $We=\rho_1 V_m^2 D_m/\sigma$ and diameter ratio of the father and mother drops, $D_r=D_f/D_m$. Here, $\rho_1,\ \rho_2$ are the densities and $\mu_1,\ \mu_2$ are the viscosities of the drop fluid and the matrix fluid, respectively. $\rho_a = (\rho_1 + \rho_1)/2$ is the average density, $\rho_c = \rho_1 - \rho_2$, is the density difference between the two fluids, and $D_m, D_f$ are the diameters of the mother and father drop, respectively. $V_m$ is the velocity of the mother drop. The length scales are non-dimensionalized using the mother drop diameter $D_m$ and the time is scaled by capillary time $\tau_c = \sqrt{\rho_a D_m^3/\sigma}$.

\subsection{Validation} \label{sec:validation}
Our numerical solver is validated by comparing the results obtained from our simulations with the earlier results for (i) the partial coalescence of a drop on a pool of the same liquid and (ii) the partial coalescence of two unequal-sized drops of the same liquid. Additionally, we have also conducted a grid convergence test for the present configuration for a typical set of parameters. 

\begin{figure}
\centering
\includegraphics[width=0.9\textwidth]{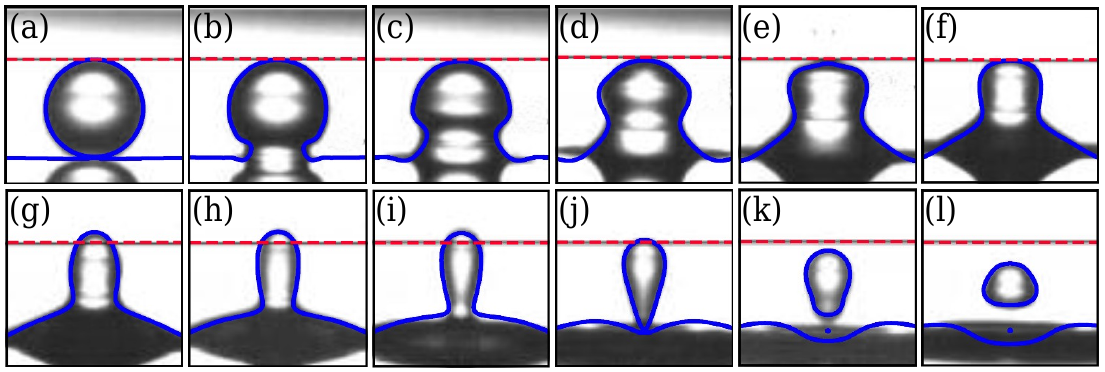}
\caption{Comparison between the present numerical results (blue line) with the experimental results (in the background) of \citet{chen2006partial} for the partial coalescence phenomenon of a drop (diameter 1.1 mm) on a liquid pool. Here, $Oh_1 = 0.0058$, $Oh_2 = 0.0117$, $Bo = 0.0958$ and $A = 0.136$. The panels are 542 $\mu$s apart in time.}
\label{fig:fig2}
\end{figure}

In figure \ref{fig:fig2}, we compare the partial coalescence of a single drop on a liquid pool obtained from our numerical simulation with the experimental result of \citet{chen2006partial}. The mechanism of partial coalescence is explained in detail in \S\ref{sec:introduction} . The capillary wave generated at the point of contact propagates upward along the drop surface and imparts an upward vertical pull which elongates the drop beyond its initial height. This can be clearly observed in figure \ref{fig:fig2}(h). The continuous drainage of liquid from the drop and the vertical pull from the capillary waves thins the liquid column. Due to the prevailing horizontal collapse rate, necking starts at the base of the column at a certain point, and a secondary drop pinches off. Figure \ref{fig:fig2} shows that our numerical results are in good agreement with the experimental results.

\begin{figure}
\centering
\includegraphics[width=0.9\textwidth]{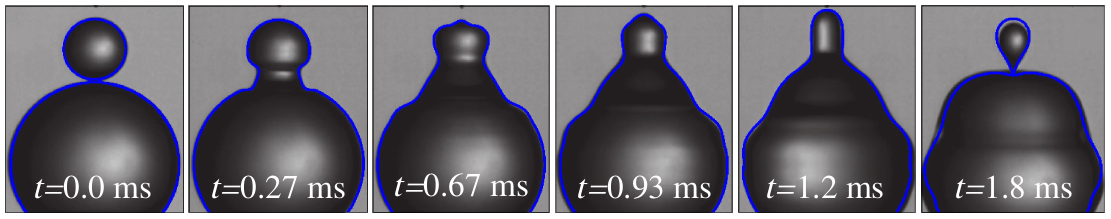}
\caption{Comparison between the present numerical results (blue line) with the experimental result of \citet{zhang2009satellite} shown as background for the partial coalescence of two unequal-sized drops with $D_r =2.72$. The other dimensionless parameters are $Oh_1 = 0.0058$, $Oh_2 = 0.000116$, $Bo = 0.092$ and $A = 0.9976$.}
\label{fig:fig3}
\end{figure}

Then we validate our solver by simulating the phenomenon of partial coalescence observed during the interaction of two vertically aligned unequal-sized drops of $D_r = 2.72$, which was experimentally investigated by \citet{zhang2009satellite}. It is observed that initially, the upper drop rests on the interface of the lower drop for a finite duration until the film between the two drops ruptures. Due to the breakup of the thin air film between the two drops, some liquid of the mother drop drains out and mixes with the father drop. After some time, it forms a columnar structure that gradually becomes thin and develops a neck at its base. Finally, the neck pinches off, generating a daughter drop. Again, the results of our numerical simulations depict a good agreement with the experimental results of \cite{zhang2009satellite}, as illustrated in figure \ref{fig:fig3}.

Further, in order to validate our solver quantitatively, we perform another simulation with $D_f/D_m = 2.87$ and compare our results with that of \citet{zhang2009satellite}. The other dimensionless parameters are $Oh_1 = 0.0082$, $Oh_2 = 0.00006$, $Bo = 0.304$ and $A = 0.9976$. The ratio of the daughter drop diameter to the mother drop diameter in our simulation is found to be 0.545 which is in good agreement with the experimentally observed value of 0.547. This agreement between the numerical and the experimental results justifies the accuracy of the simulations.

\subsection{Grid independence}
\begin{figure}
\centering
\includegraphics[width=0.95\textwidth]{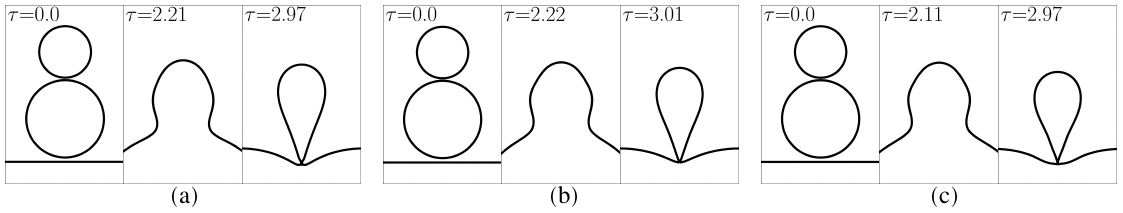}
\caption{Coalescence dynamics of two unequal-sized drops ($D_r=1.5$) and a pool obtained using grid sizes (a) $D_m/65$, (b) $D_m/82$ and (c) $D_m/100$. Other dimensionless parameters are $Oh_1 = 0.0058$, $Oh_2 = 0.000116$, $Bo = 0.092$, $A = 0.9976$ and $We = 0.11$.}
\label{fig:fig4}
\end{figure}

In order to ensure the grid independence, we perform numerical simulations using three grids sizes for the case with $D_r = 1.5$ in the configuration shown in figure \ref{fig:fig1}. The results obtained using the grid sizes $D_m/65$, $D_m/82$ and $D_m/100$ are shown in figure \ref{fig:fig4}(a)-(c). The values of the corresponding dimensionless radius of the resultant satellite drop are 0.526, 0.508 and 0.505, respectively. The percentage difference between the values of radii for grid sizes $D_m/82$ and $D_m/100$ is approximately 0.72 \%. Thus, the intermediate grid size of $D_m/82$ has been considered for the rest of the simulations.

\section{Results and Discussion}\label{sec:results}
\subsection{Coalescence cascade of the conglomerate}
The diameter ratio between the father and mother drops ($D_r$) determines whether the two drops coalesce completely or partially \citep{zhang2009satellite, deka2019coalescence}. For water drops in the surrounding air, with the dimensionless parameters, $Oh_1 = 0.0058$, $Oh_2 = 0.000116$, $Bo = 0.092$ and $A = 0.9976$ (kept fixed in our study), we have performed simulations for different $D_r$ values. Unless mentioned explicitly, throughout the study, $We=0.11$. The diameter ratio is varied from 1.0 to 3.0, keeping the size of the mother drop constant. This range encompasses the transition from total to partial coalescence for two drops while the interaction of the conglomerate with the liquid pool always produces partial coalescence. Although, at higher $We$, the conglomerate also shows a transition from total to partial coalescence during its interaction with the pool as the value of $D_r$ is increased.

\begin{figure}[H]
    \centering
    \includegraphics[width=0.9\textwidth]{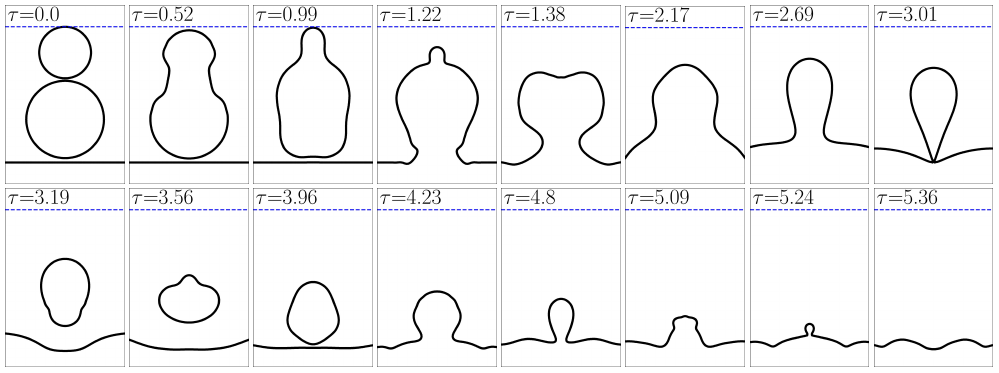}
    \caption{Coalescence dynamics of two unequal-sized drops ($D_r=1.5$) and a pool for $We = 0.11$. The blue dashed lines represent the initial locations of the tips of the mother drop.}
    \label{fig:fig5}
\end{figure}

\begin{figure}[H]
    \centering
    \includegraphics[width=0.9\textwidth]{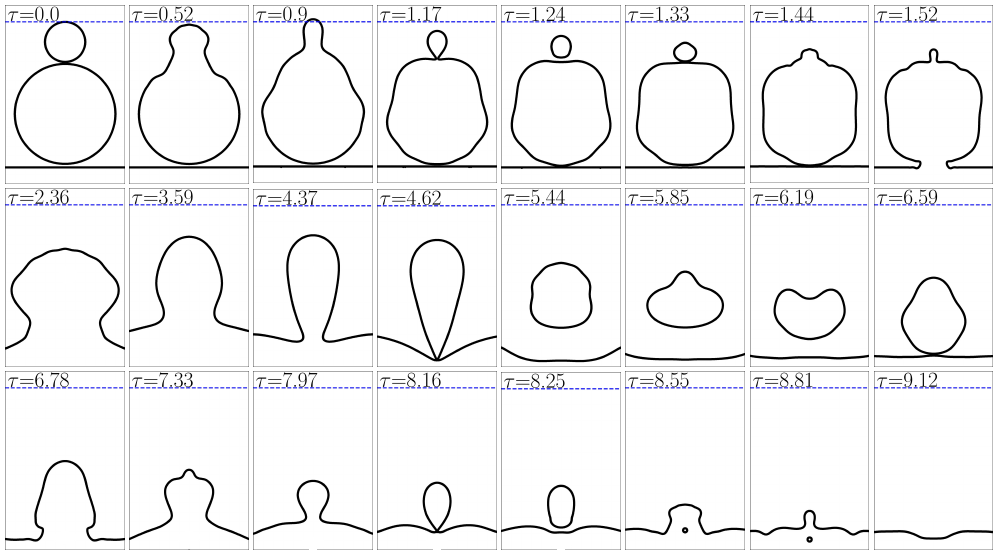}
    \caption{Coalescence cascade for $D_r= 2.5$ and $We = 0.11$.}
    \label{fig:fig6}
\end{figure}

We observe that the two drops merge fully for $D_r \le 2.25$. For higher $D_r$ values, the mother drop coalesces partially with the father drop and produces a single-stage cascade. For $D_r =1.5$, the mother drop coalesces completely with the father drop, as shown in figure \ref{fig:fig5}. When the conglomerate is deposited onto the pool, it produces a satellite drop, which eventually coalesces with the pool. Figure \ref{fig:fig6} shows another situation where the two drops coalesce partially, producing a satellite (daughter) drop. Here, the conglomerate exhibits a two-stage cascade with the pool. Among the $D_r$ values considered here, the conglomerate produces a single-stage cascade with the pool for $D_r = 1$, 1.25 and 1.5; while for higher $D_r$ values, it produces a two-stage cascade.

\ks{It is to be noted that \citet{tang2012} observed that the critical Weber number, which separates bouncing and permanent coalescing regimes, does not vary considerably with the diameter ratio of two droplets undergoing head-on collision. In our study, although we did not observe bouncing behavior for vertically aligned drops, we noticed that some conglomerate morphologies bear a striking resemblance to the coalescence regime reported by \citet{tang2012}.}

\subsection{Effect of pool on the coalescence dynamics of two drops}\label{res:b}

\begin{figure}[h]
    \centering    
    \includegraphics[width=0.8\textwidth]{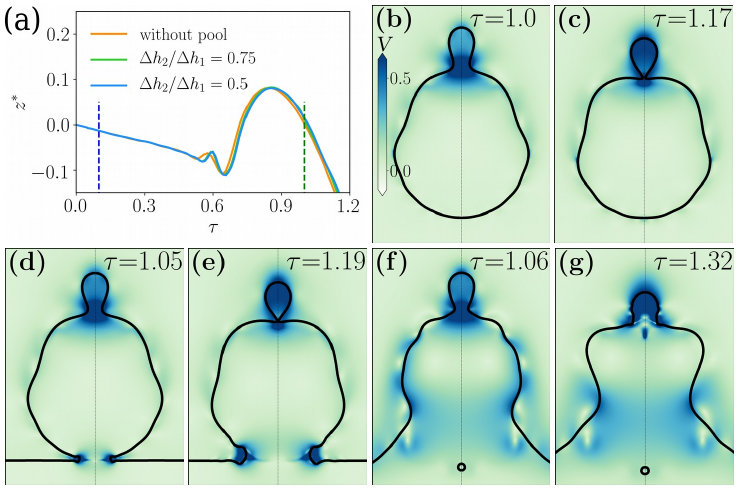}
    \vspace{-3mm}
    \caption{Temporal variation of the dimensionless location of the tip of the mother drop $(z^*=z-z_0/D_m,\ z_0$=initial height of the tip of the mother drop) in the presence and absence of the pool. The blue and green vertical dashed lines in (a) denote the instants when the father drop starts draining into the pool for $\Delta h_2/ \Delta h_1 = 0.5$ and 0.75, respectively. The contour represents the resultant velocity ($V$) for $D_r= 2.5$ and $We = 0.11$. Panels (b) and (c) represent two sequences during father-mother drop interaction when the pool is absent, while panels (d-e) and (f-g) show the sequences when $\Delta h_2/ \Delta h_1 = 0.75$ and 0.5 respectively. }
    \label{fig:fig7}
\end{figure}

In order to estimate the effect of the presence of a pool on the coalescence dynamics of the two drops, we performed a set of simulations for $D_r$=2.5. In this analysis, the value of $\Delta h_2/\Delta h_1$ is suitably adjusted to ensure that the two drops and the father drop and the pool start coalescing simultaneously. We allowed the mother and father drop to coalesce in the absence of pool for one case while for the other two cases, $\Delta h_2/\Delta h_1$ is varied from 0.5 to 0.75. Figure \ref{fig:fig7}(b) and (c) refer to the case where the pool is absent, and it can be seen that a daughter drop is pinched off from the interaction of the mother and father drops. Figure \ref{fig:fig7}(d,e) shows the case for $\Delta h_2/\Delta h_1 = 0.75$. Here, we can see that the daughter drop pinches off in spite of the interaction between the conglomerate and the pool. As we further decrease $\Delta h_2/\Delta h_1$ to 0.5, we can see from figure \ref{fig:fig7}(f,g) that the pinch-off of the daughter drop is prevented due to excessive drainage of the conglomerate into the pool. The vertical collapse rate of the liquid column expedites as a result of this drainage, which also causes the mother drop to start draining more into the father drop (figure \ref{fig:fig7}(f)). It is well known that the competition between the horizontal and vertical rates of collapse leads to partial coalescence. In this scenario, the vertical rate of collapse dominates over the horizontal rate of collapse and partial coalescence is prevented. However, in all these cases it is observed from figure \ref{fig:fig7}(a) that the apex of the mother reaches the same level of height irrespective of the position of the pool.

\subsection{Capillary waves}\label{sec:waves}

\begin{figure}
    \centering
    \includegraphics[width=0.95\textwidth]{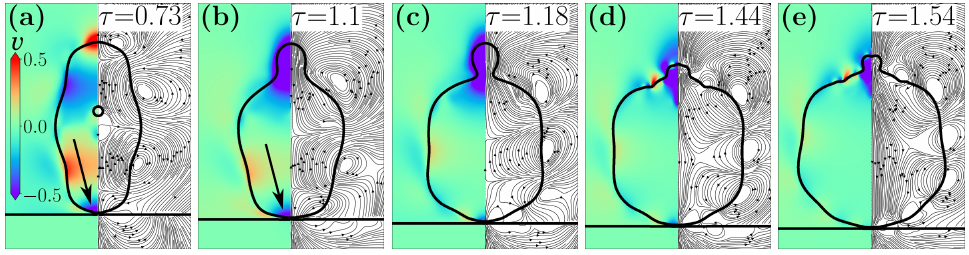}
    \caption{The axial velocity $(v)$ field and streamline pattern at the onset of coalescence of the conglomerate with the pool at $We$ = 0.11 for (a) $D_r = 1$ (b) $D_r = 1.5$ (c) $D_r = 2$ (d) $D_r = 2.5$ and (e) $D_r = 3$.}
    \label{fig:fig8}
\end{figure}

The capillary waves are generated by the interplay between surface tension and the pressure variation in the liquid. These waves, with shorter wavelengths, carry enough momentum to influence the coalescence characteristics of the drops. In the case of a drop interacting with a pool of the same liquid, \citet{blanchette2006partial} reported that the partial coalescence depends on the strength of these capillary waves to pull the drop vertically from the top to stretch it. During drop-drop or drop-pool interaction, the capillary waves generated at the interface travel along the interacting fluids. The effect of the capillary waves can be observed profoundly in the vertical stretching of the impinging drop. In figure \ref{fig:fig6}, at $\tau= 0.9$, it can be noticed that the mother drop is elongated beyond the blue dashed line.

In the simple configurations of a drop interacting with a pool or another drop, the coalescence dynamics can be anticipated to a reasonable extent by deciphering the paths of the capillary waves. However, in a complex configuration, such as the one considered in this study, the propagation and interaction of the capillary waves generated at multiple interfaces can influence the coalescence process in unpredictable ways. During the coalescence of a drop on a liquid pool, the capillary waves that travel along the liquid pool do not affect the coalescence dynamics, but in the coalescence of a pair of drops, the capillary waves travelling along the father drop sometimes converge at the bottom and impart a downward pull on the father drop. Also, when the father drop interacts with the pool, another set of capillary waves is generated, some of which start moving upwards along the periphery of the father drop. The effect of these waves on the coalescence dynamics will be discussed in the context of the interaction of the conglomerate with the pool.

When the two drops coalesce, a conglomerate is formed. In the initial stages, the mother drop contributes to the upper part, while the father drop contributes to the bottom part of the conglomerate. The capillary waves travel in both directions from the location where the mother and father drops interact. The movement of the capillary waves affects the morphology of the conglomerate during coalescence. Figure \ref{fig:fig8} illustrates the shape of the conglomerate, vertical velocity field and streamline pattern at the onset of coalescence with the pool for different $D_r$ values. As the diameter ratio increases, the capillary waves cease to significantly deform the bottom part of the conglomerate, and it assumes a near-spherical shape till the onset of its coalescence with the pool. For $D_r = 1$, the capillary waves travelling along the periphery of the father drop converge at the bottom of the father drop. This happens before the conglomerate can start coalescing with the pool, and it is evident from the considerable downward momentum in the bottom part of the conglomerate (figure \ref{fig:fig8}(a)). A similar downward momentum can also be seen in figure \ref{fig:fig8}(b) for $D_r = 1.5$. However, with the increase in the size of the father drop, the downward momentum as shown by black arrows in figure \ref{fig:fig8}(a) and (b) gradually vanishes. This result indicates that for small $\Delta h_2$ values ($\Delta h_2/\Delta h_1 = 1.75$), the downward momentum developed owing to the converging capillary waves facilitates the coalescence process. On the other hand, for higher $D_r$ values, the capillary waves travelling along the mother drop converge earlier at the top, imparting an upward pull on the conglomerate and delaying the coalescence process.

\begin{figure}
    \centering
    \includegraphics[width=0.9\textwidth]{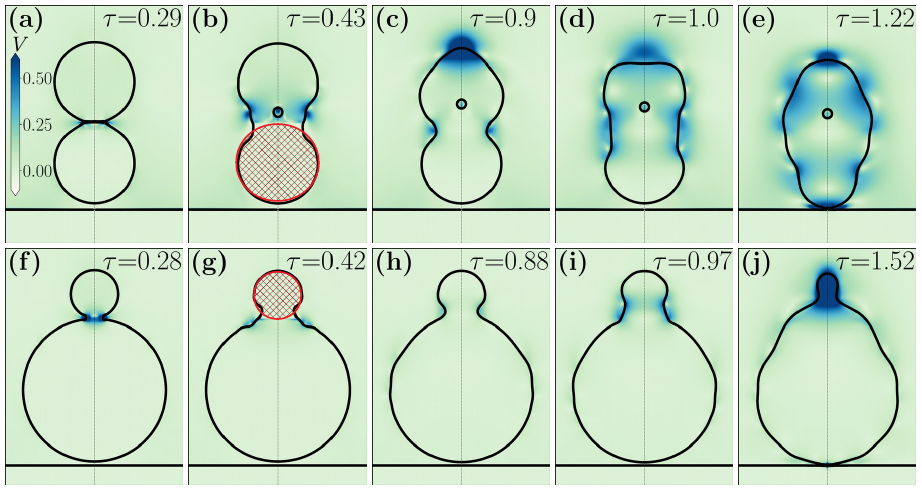}
    \caption{Evolution of the conglomerate during the analysis of the effect of capillary waves by turning the velocity field off for 0.2 $ms$ in the father drop for $D_r = 1$ (a)-(e) and in the mother drop for $D_r$ = 3 (f)-(j). The cross-hatched parts in (b) and (g) represent the areas where the velocity is turned off. $We$ value is 0.11.}
    \label{fig:fig9}
\end{figure}

To verify the effect of the capillary waves in facilitating/delaying the coalescence of the conglomerates for lower/higher diameter ratios, we perform tests similar to that of \cite{blanchette2006partial} and \cite{deka2019coalescence} by setting the velocity to zero after a particular time instant and restarting the simulation. Here, we interrupt the velocity field for the extreme $D_r$ values in our range. The cross-hatched regions in figure \ref{fig:fig9}(b) and (g) identify the regions where the velocity field is turned off. The upper row (figure \ref{fig:fig9}(a)-(e)) represent the evolution of the conglomerate for $D_r = 1$ while the bottom row (figure \ref{fig:fig9}(f)-(j)) represent the case for $D_r = 3$. We set the velocity to zero in the lower half of the conglomerate for $D_r = 1$, for 0.2 ms at $\tau = 0.43$, and the velocity field distributes naturally after that. By doing this, the effect of the capillary waves in facilitating the coalescence between the conglomerate and the pool is reduced. This is evident from the time instant at which the conglomerate starts coalescing with the pool, which is delayed ($\tau = 1.22$) compared to the case ($\tau = 0.73$) when the velocity field was not interrupted. A similar study was conducted for $D_r = 3$, where the velocity field was turned off in the upper part of the conglomerate just after the two drops started coalescing with each other. Due to the interruption of the velocity field, the upward pull exerted by the converged capillary waves on the top of the conglomerate is expected to decrease. This indeed advanced the coalescence of the conglomerate with the pool from $\tau=1.54$ (figure \ref{fig:fig8}(e); original case of uninterrupted velocity field) to $\tau=1.52$, as shown in figure \ref{fig:fig9}(j). Also, due to the weakening of the upward pull on the top part of the conglomerate, the daughter droplet generation is inhibited in this case.

\begin{figure}
    \centering
    \includegraphics[width=0.99\textwidth]{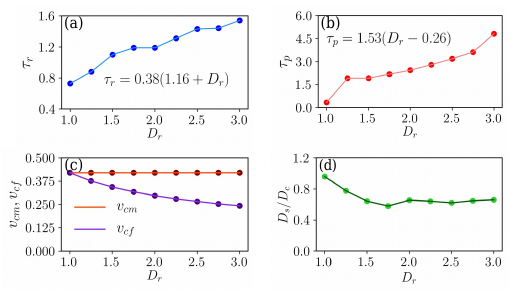} \\
    \vspace{-5.0 mm}
    \caption{The variations of (a) $\tau_r$ (b) $\tau_p$  (c) $v_{cm}$ \& $v_{cf}$ and (d) $D_s/D_c$ with $D_r$ for $We = 0.11$.}
    \label{fig:fig10}
\end{figure}

The time duration for the bottom-most point of the father drop to reach from its initial position to a position at the onset of coalescence of the drop-conglomerate with the pool is defined as residence time. \ks{The variation of the dimensionless residence time ($\tau_r$)} with $D_r$ is plotted in figure \ref{fig:fig10}(a) and it shows that $\tau_r$ increases continuously as $D_r$ is increased. As discussed earlier, the capillary waves travelling along the interface of the father drop in the downward direction would impart a downward pull if converged at the bottom of the conglomerate. At the same time, the capillary waves, moving along the interface of the mother drop, try to pull the drop upwards. With the increasing diameter ratio, the capillary waves travelling downwards along the father drop become ineffective in imparting the downward pull as they cannot travel to the bottom of the conglomerate. On the contrary, due to the increased flatness of the father drop and higher curvature at the point of contact of the two drops, the capillary waves impart a stronger upward pull. This combined effect contributes to an increase in $\tau_r$ with the increase in diameter ratio. 

Furthermore, increasing $D_r$ increases the \ks{dimensionless} pinch-off time ($\tau_p$), which can be defined as the time duration from the onset of coalescence to the pinch-off of the first satellite during the conglomerate-pool interaction. This is because, as the size of the father drop increases, the capillary waves require more time to climb to the top of the conglomerate and impart the vertical pull, which eventually contributes significantly to the pinch-off process. Figure \ref{fig:fig10}(b) shows that increasing $D_r$ increases $\tau_p$. However, in addition to the size, we noticed that the shape of the conglomerate also plays an important role in determining the pinch-off time. In contrast to higher $D_r$ values, the pinch-off time for $D_r = 1$ is quite shorter ($\tau=0.34$). \ks{We observe that the capillary pressure within the drops, the momentum carried by the generated capillary waves, and the local curvature of the neck all along affect the pinch-off dynamics to set the critical limit of pinch-off time. A decrease in the size of the father drop increases the capillary pressure within the father drop, which increases the radial expansion of the neck after coalescence leading to an increase in drainage. Because of the smaller size of the father drop, the curvature in the neck region becomes less sharp. As a result, the capillary waves produced because of the curvature change near the contact region become weaker and carry less momentum. This decreases the vertical stretching of the mother drop and affects the pinch-off dynamics. Moreover, because of less stretching of the mother drop, the local curvature at the neck region increases in the retraction stage of the neck, which also affects the pinch-off dynamics. The rate of the downward pull of the liquid is controlled by the capillary speed of the mother drop, while the capillary speed of the father drop controls the upward pull. The capillary speeds of the mother drop ($v_{cm}$) and father drop ($v_{cf}$) can be defined as
\begin{eqnarray}
    v_{cm} &=& D_m \sqrt{\frac{\sigma}{\rho_a D_m^3}},  ~ {\rm and} \\
    v_{cf} &=& D_f \sqrt{\frac{\sigma}{\rho_a D_f^3}}.
\end{eqnarray}
Figure \ref{fig:fig10}(c) shows the relationship between $D_r$ and the variations of $v_{cm}$ and $v_{cf}$. The capillary speed of the mother drop, which has a fixed diameter of $0.82 \ {\rm mm}$, remains constant at $0.42 \ {\rm ms}^{-1}$. However, the capillary speed of the father drop decreases as $D_r$ increases due to its larger diameter. As depicted in figure \ref{fig:fig10}(c), this decrease in capillary speed causes the capillary waves on the father drop to take longer to converge at the bottom, facilitating coalescence. Our simulation shows that the residence time ($\tau_r$) is related to the diameter ratio as $\tau_r=0.38(1.16+D_r)$, and the pinch-off time ($\tau_p$) is related to $D_r$ as $\tau_p=1.53(D_r-0.26)$.
}

\ks{Figure \ref{fig:fig10}(d)} shows the variation of $D_s/D_c$ with diameter ratio. Here, $D_s$ is the equivalent diameter of the satellite, and $D_c$ is the equivalent diameter of the conglomerate. For $D_r=1$, it is seen that the satellite drop is of almost the same size as that of the conglomerate ($D_s/D_c=0.959$), indicating that there is very less drainage in the first stage of coalescence. The ratio gradually decreases with the increase in $D_r$ and finally settles to a nearly constant value. The plausible reasons for the smaller value of $\tau_p$ and higher $D_s/D_c$ for smaller values of $D_r$ is explained in detail in the next section.

\subsection{Pinch-off dynamics of the conglomerate of equal-sized drops}

\begin{figure}
    \centering
    \includegraphics[width=0.95\textwidth]{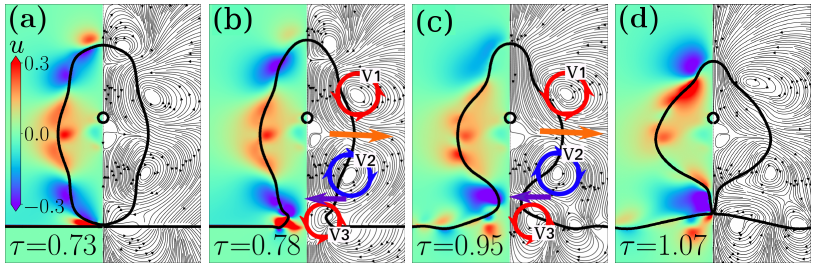}
    \caption{Pinch-off of the satellite during the interaction of the conglomerate with the pool for $D_r = 1$ and $We = 0.11$. The left half and right half of the figire show the radial ($u$) velocity and the streamlines, respectively.}
    \label{fig:fig11}
\end{figure}

In figure \ref{fig:fig11}, we have shown the evolution of the conglomerate and the pinch-off of the satellite during the conglomerate-pool interaction for $D_r = 1$. The minimal $\tau_p$ value for this case indicates that the pinch-off occurs quite rapidly.  The non-columnar shape at pinch-off can be seen in figure \ref{fig:fig11}(d) at $\tau = 1.07$. The shape of the conglomerate in this case resembles a prolate drop. \citet{biswaseffect} studied the effect of drop shape on partial coalescence and reported that for the prolate and oblate-shaped drops, the necking and pinch-off phenomena happen faster as compared to a spherical drop. However, in this situation, the effect of shape is complemented by the complex interaction of capillary waves as we will see next. We find that the drop may not always assume a conventional columnar shape during the pinch-off. During the satellite pinch-off, similar shapes (like figure \ref{fig:fig11}(d)) have been observed by \citet{ray2013clsvof} when two drops were allowed to impinge on a pool within a small time interval (10 ms). \citet{yi2014temperature} had similar observations during the head-on collision of binary droplets on a superhydrophobic surface.

The narrowing of the top and the bottom part of the structure is due to the negative radial $(u)$ velocity, clearly seen in the bluish regions of the contour in figure \ref{fig:fig11}(a-d), while the outward radial bulge in the middle part of the satellite is developed due to the positive $u$-velocity shown by the reddish region in the middle. The velocity field that engenders this unconventional structure during pinch-off can be attributed to the interplay of the capillary waves generated at multiple liquid-liquid junctions. From the streamlines of figure \ref{fig:fig11}(b) and (c), it can be seen that the capillary waves generated during the interaction of the two drops have an opposite sense of rotation. The waves travelling along the upper and the lower drops rotate in anti-clockwise and clockwise directions and are shown by a red counterclockwise rotating circle ($V_1$) and blue clockwise rotating circle ($V_2$), respectively. This pair of capillary waves impart an outward momentum in the central part of the fluid structure, which is shown by the orange arrows. While travelling down the conglomerate, the clockwise rotating capillary wave ($V_2$) encounters a counterclockwise rotating capillary wave near the neck region ($V_3$), which is generated at the conglomerate-pool interface. These two capillary waves direct the flow in such a way that it achieves sufficient inward momentum near the neck region to cause an early pinch-off of the satellite.

The early pinch-off effectively stops the conglomerate from draining into the pool, and the resultant satellite retains much of the fluid mass of the conglomerate. Hence, this study clearly explains  higher $D_s/D_c$ values associated with the smaller $D_r$ values. 

\subsection{Comparison with that of a spherical drop}\label{sec:spherical}

Here, for each diameter ratio, we compare the coalescence dynamics of the conglomerate with that of an equivalent spherical drop of the same volume. Figures \ref{fig:fig12}(a)-(f) illustrate the interface profiles at the onset of coalescence (top) and during satellite pinch-off (bottom) for the conglomerate and the spherical drop for different $D_r$ values. Due to no prior interactions, the spherical drop is devoid of any capillary waves, unlike that of the conglomerate. Figures \ref{fig:fig12}(a) and (b) refer to the case of the conglomerate and the spherical drop respectively. Similarly, figures \ref{fig:fig12}(c) and (d) represent the cases for $D_r$ = 1.5 and figures \ref{fig:fig12}(e) and (f) represent the cases for $D_R$ = 2.

\begin{figure}[H]
\centering
\includegraphics[width=0.9\textwidth]{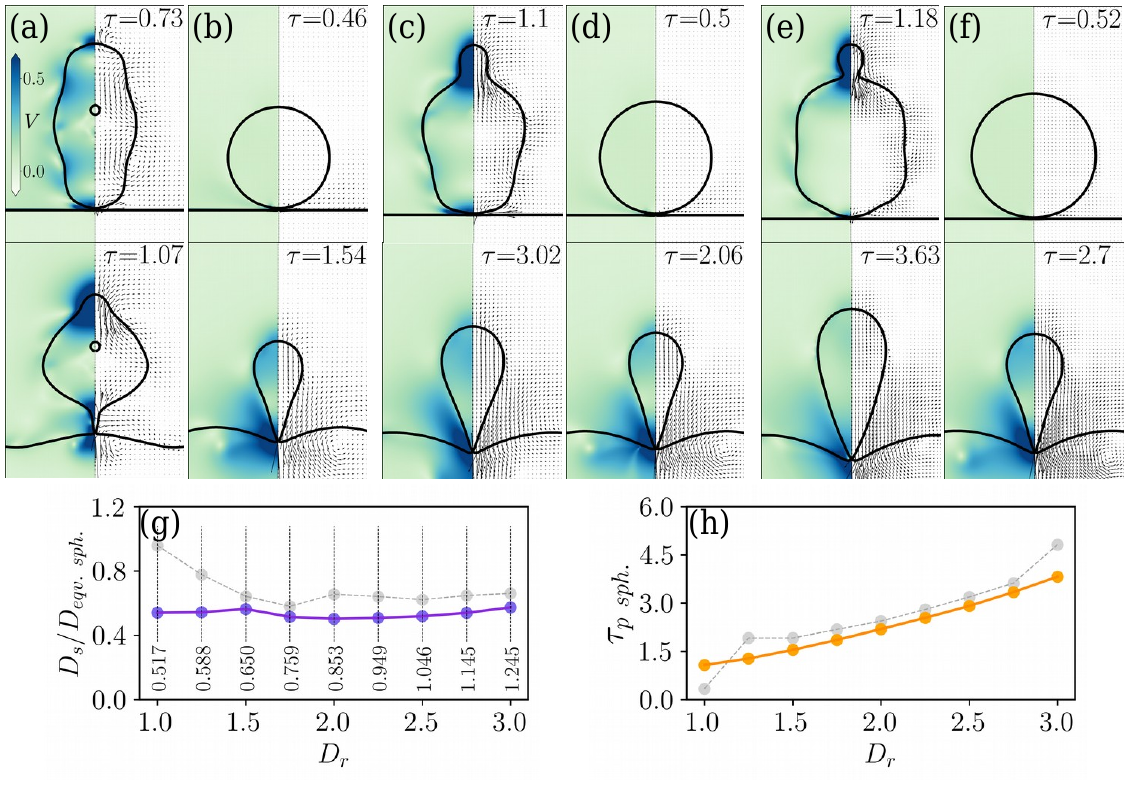}\label{fig:fig12}
\vspace{-3.0mm}
\caption{Panels (a) and (b) show the comparison of interface profiles at the onset of coalescence (top) and during pinch-off (bottom) for the conglomerate and the spherical drop respectively, for $D_r = 1$. Similarly, panels (c) and (d) and panels (e) and (f) show the comparison for $D_r$ =  1.5 and 2, respectively. Figure (g) shows the variation of $D_s$/$D_{eqv. \ sph.}$ with $D_r$, while (h) shows the variation of $\tau_{p \ sph.}$ with $D_r$ for the equivalent spherical drop. The variations of $D_s/D_c$ and $\tau_p$ for the conglomerate have been shown using gray dashed curves in the background. The radius of the equivalent spherical drop corresponding to each $D_r$ value is mentioned inside panel (g). $We$ value is 0.11.}
\label{fig:fig12}
\end{figure}

It can be seen that the interface profiles of the conglomerate and the spherical drop vary the most for $D_r = 1$. As $D_r$ increases, the profiles shown in figure \ref{fig:fig12}(c)-(f) indicate that the shape of the conglomerate at the onset of coalescence approaches towards a spherical form as the diameter ratio is increased (more clearly observed in figure \ref{fig:fig8}). Also, the columnar structure during satellite pinch-off for the conglomerate and the spherical drop become almost indistinguishable as we increase $D_r$.

Figure \ref{fig:fig12}(g) shows the variation of the ratio of the diameter of the satellite ($D_s$) and the equivalent spherical drop ($D_{eqv. \ sph.}$) corresponding to each $D_r$. For higher $D_r$ values, the pinch-off dynamics of the conglomerate and the spherical drop shows qualitatively similar behaviour. However, it is clearly seen that for smaller diameter ratios, the size of the satellite produced during the interaction of a purely spherical drop is much smaller than the satellite produced during the conglomerate-pool interaction. For the spherical drop, the $D_{s}/D_{eqv. \ sph.}$ ratio remains nearly constant irrespective of the size of the impinging drop, while for the conglomerate, we see that due to the change in the shape of the impinging liquid mass and complex interaction of capillary waves (described in \S\ref{sec:waves}), the ratio starts-off with a high value and gradually decreases. Here, $\tau_{p \ sph.}$ is the pinch-off time for the satellite produced during the interaction of the equivalent spherical drop. $\tau_{p \ sph.}$ increases consistently as $D_r$ is increased unlike that in case of the conglomerate where a sudden jump was observed as $D_r$ was increased from 1 to 1.25 (figure \ref{fig:fig12}(h)). This further demonstrates that as $D_r$ increases, the effect of shape reduces and the coalescence dynamics of the conglomerate approaches that of a pure spherical drop.


\subsection{Comparison with a drop with a conglomerate-like initial shape}\label{res:f}

\begin{figure}
\centering
\includegraphics[width=0.9\textwidth]{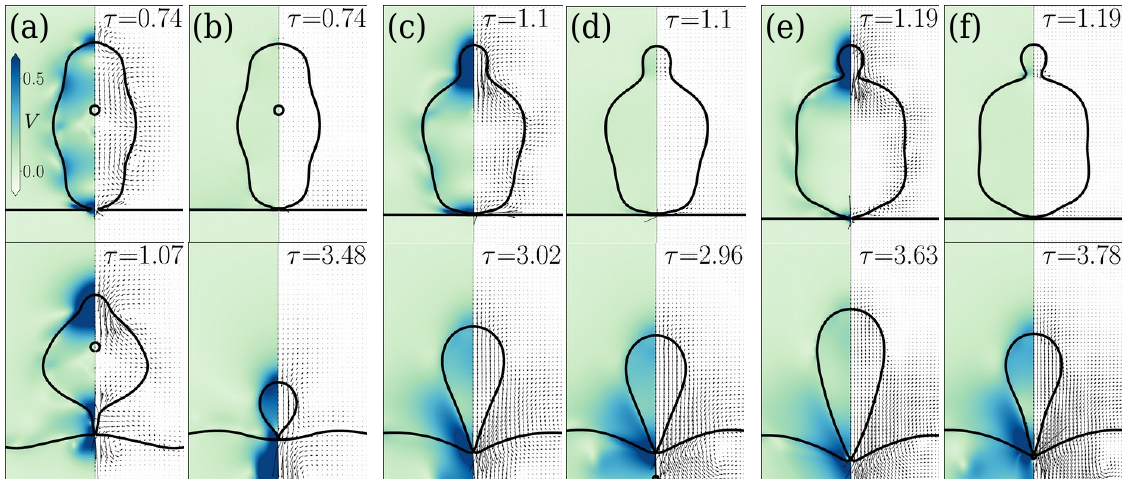}
\vspace{-3.0mm}
\caption{The panels (a), (c) and (e) show the comparison of coalescence dynamics of a conglomerate for $D_r$ = 1, 1.5 and 2 respectively. Panels (b), (d) and (f) correspond to a drop initialized with the same shape as that of the conglomerate for $D_r$ = 1, 1.5 and 2, respectively. $We$ value is 0.11.}
\label{fig:fig13}
\end{figure}

In \S\ref{sec:spherical}, we have seen that the coalescence of a conglomerate in a pool differs from that of the spherical drop. In this section, we compare the coalescence-dynamics of the conglomerate with a drop of the same shape as that of the conglomerate. This drop can be termed as `conglomerate-shaped’ drop. The conglomerate-shaped drop possesses the shape of the conglomerate but it does not have the presence of the capillary waves generated during the interaction of the father and the mother drop.

Figure \ref{fig:fig13} compares the instants of the onset of coalescence (top) and pinch-off of the secondary drops (bottom) during the coalescence of the conglomerate and the conglomerate-shaped drop for various diameter ratios. Figures \ref{fig:fig13}(a) and (b) correspond to the cases of conglomerate and the conglomerate-shaped drops respectively, for $D_r$ = 1. Similarly, figures \ref{fig:fig13}(c) and (d) and figures \ref{fig:fig13}(e) and (f) correspond to the cases for the conglomerate and the conglomerate-shaped drop for $D_r$ = 1.5 and 2 respectively.


For $D_r = 1$, we still do not observe the early pinch-off of the secondary drop which was seen when the conglomerate interacted with the pool, clearly indicating that the capillary waves produced during the drop-drop interaction played a major role for the early pinch-off of the secondary droplet.

When a drop is allowed to impact on a stationary drop placed on a super-hydrophobic surface, it is observed that during the retraction phase, the coalesced droplet does not retain any memory owing to its drop-drop interaction and behaves as if it is an isolated primary drop \citep{damak2018expansion}. However, in our study, where we employ a similar configuration (the hydrophobic surface being replaced by a pool of the same liquid as that of the drop), the conglomerate is found to retain the memory of its origin in terms of the capillary waves generated during the interaction of the father drop with the mother drop. For smaller $D_r$ values these capillary waves interact with the capillary waves generated during the interaction of the conglomerate and the pool. This facilitates early pinch-off of the secondary droplet (figures \ref{fig:fig13}(a) and (b)).
However, for the larger diameter ratios, the capillary waves do not interact and an early pinch-off does not occur. From figures \ref{fig:fig13}(e) and (f), it can be observed that the coalescence-dynamics of the conglomerate and the conglomerate-shaped drop become quite similar.

\subsection{Effect of Weber number}\label{res:g}

\begin{figure}
\centering
\includegraphics[width=0.7\textwidth]{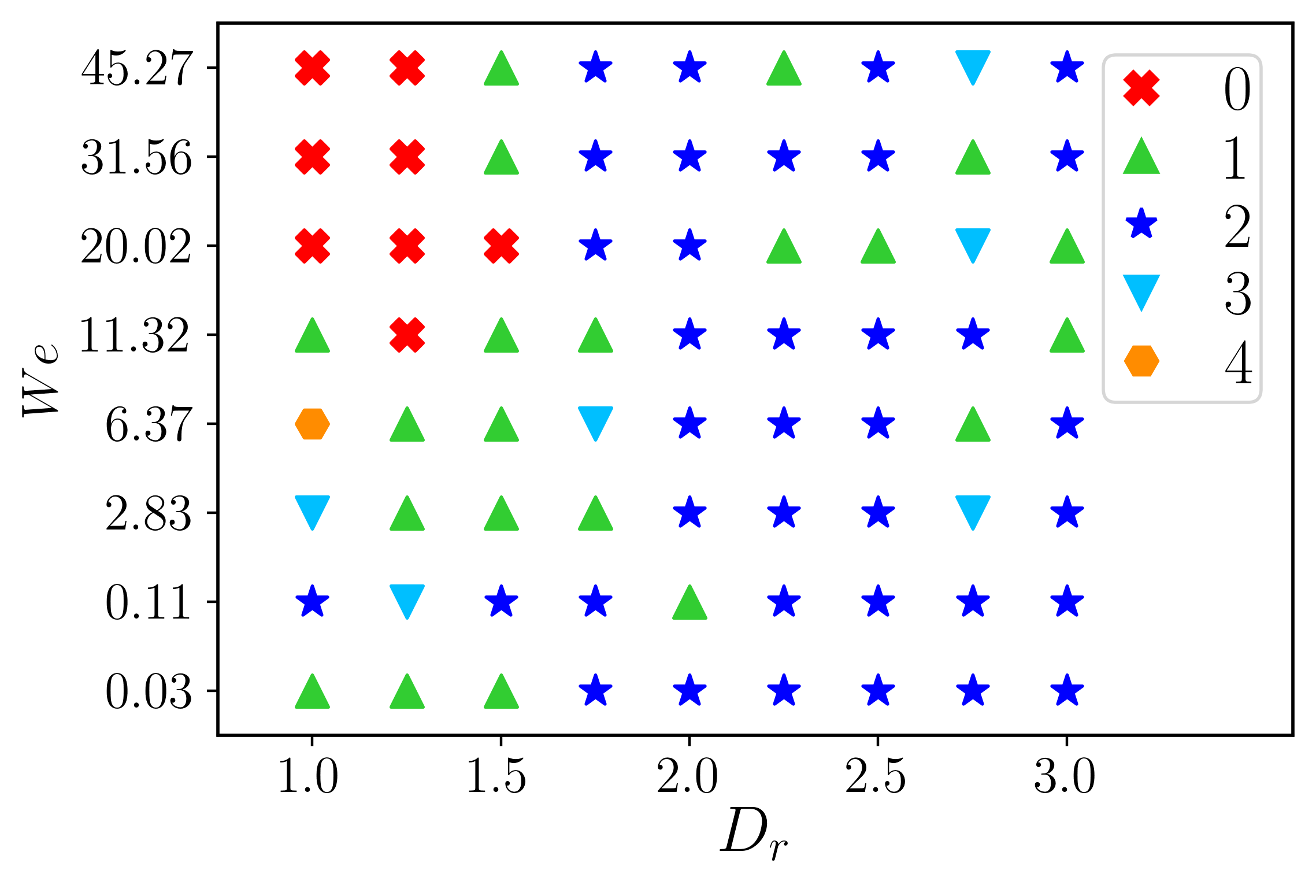}
\vspace{-5.0mm}
\caption{Regime map demarcating the complete and partial coalescence behaviours in $We$ and $D_r$ space for conglomerate-pool interaction. The numbers in the legend represent the stages of the cascade. The red-crosses corresponding to zero stages of cascade represent complete coalescence.}
\label{fig:fig14}
\end{figure}

With the variation of Weber number ($We$), we observe different coalescence patterns in drop-drop and conglomerate-pool interactions. In figure \ref{fig:fig14}, we plot the regime map that identifies different coalescence characters of the conglomerates for various $D_f/ D_m$ ratio. We observe non-monotonic behavior related to the number of satellite drops with increasing $We$. In the case of smallest diameter ratio, we observe one satellite formation for $We=0.03$. While for $We=0.11$, we observe the formation of two satellites.  As we increase the $We$ to 6.37, four satellites are formed. Seemingly, in this range of $We$, it is a monotonic increase in the number of satellites. The number of satellites reduces to one for $We=11.32$. Subsequently, for the higher Weber numbers, for example, for $We=20.02$, the conglomerate coalesces completely without forming any satellite, and $We=31.56$ repeats the complete coalescence behavior. However, for $We=45.27$, we observe complete coalescence with a tendency for a jet formation at the bottom surface of the crater. As we increase $D_r$, we expect to see complete coalescence at even higher $We$ values but at $D_r = 1.25$, we surprisingly notice that complete coalescence starts at a lower $We$ of 11.32 compared to $We = 20.02$ in case of $D_r = 1$. The cascading process for the $D_r = 1.25$ starts off by producing a single satellite at $We = 0.03$, increasing to three satellites at $We = 0.11$. The number of satellites produced reduces again to one for higher $We$ and remains constant till the start of complete coalescence at $We = 11.32$. For $D_r = 1.5$, the whole range of $We$ spans through partial coalescence except for an intermediate value of $We = 20.02$. Here too, we see a single satellite formation at the start, followed by the formation of two satellites at $We= 0.11$, and the number of satellites produced gets reduced again as $We$ is increased. For higher $D_r$ values, the regime of complete coalescence completely vanishes, and partial coalescence of various stages of cascade prevails throughout the $We-D_r$ space. From the regime map, we can see that the top-left portion represents the regions of complete coalescence suggesting that lower diameter ratios and higher Weber numbers may lead to complete coalescence. However, the number of satellites produced throughout this $We-D_r$ space varies arbitrarily, indicating the strong influence of the shape of the conglomerate and the complex interaction of capillary waves in determining the overall coalescence behaviour.

\section{Conclusions}\label{sec:conclusion}

The present study enables one to understand the coalescence dynamics of an arbitrary-shaped liquid conglomerate with a pool at a low impact velocity. The investigation is conducted through numerical simulations using an in-house flow solver. The effect of the diameter ratio ($D_r$) and the Weber number $(We)$ on the coalescence dynamics is examined by keeping the size of the mother drop fixed. For a low value of Weber number ($We = 0.11$), we observe a transition from complete to partial coalescence for two interacting drops as $D_r$ increases. However, the liquid conglomerate exhibits a partial coalescence behaviour with the pool for all the values of $D_r$ investigated. The influence of the pool on the coalescence dynamics of the two drops is also quantified. We demonstrate that even though the extent to which the apex of the mother drop is stretched during its coalescence with the father drop is nearly the same for the cases with and without the presence of the pool, satellite generation is inhibited for cases where the pool is too close to the father drop. The time duration for the bottom-most point of the father drop to reach from its initial position to a position at the onset of coalescence of the conglomerate with the pool ($\tau_r$) is influenced by the movement of the capillary waves generated during the interaction between the father and mother drops. The pinch-off time ($\tau_p$) for the first satellite drop increases consistently with an increase in $D_r$ when the conglomerate and pool are considered. The pinch-off time for the first satellite drop becomes minimal for $D_r = 1$. This is explained in the context of an interplay between the shape of the conglomerate and the complex interaction of the capillary waves. A new morphology has also been observed during the pinch-off of the first satellite drop for $D_r = 1$. The morphology looks different from the columnar structure usually observed during a normal pinch-off. The variation of the ratio of the size of the satellite to that of the conglomerate ($D_s/D_c$) with $D_r$ is studied. It is found that decreasing the value of $D_r$ increases the value of $D_s/D_c$. The previously measured parameters ($\tau_p$ and $D_s/D_c$) for the conglomerate-pool interaction are then compared to the cases involving purely spherical drops, and the deviations are reported. Additionally, the dynamics of a falling drop that is initiated with the shape of the conglomerate is compared to that of the conglomerate. We observe that, in contrast to drops coalescing over superhydrophobic surfaces, the coalesced conglomerate retains the memory of its origin. Finally, a regime map on a $We-D_r$ space is constructed, which demarcates the complete and partial coalescence regions for the conglomerate-pool interaction. The arbitrary variation in the number of satellites produced throughout this $We-D_r$ space shows the influence of the shape of the conglomerate and complex trajectories of capillary waves in dictating the general coalescence process. Hence, the current investigation opens up the scope of further studies unravelling the effect of drop shape in coalescence dynamics.

\section{Acknowledgements:} G.B. acknowledges his gratitude to J. C. Bose National Fellowship of SERB, Government of India (JBR/ 2021/ 000042). K.C.S. thanks the Science \& Engineering Research Board, India for the financial support through grant CRG/2020/000507.


\end{document}